\documentclass[prb,showpacs,amsmath,amssymb]{revtex4}

\usepackage{graphicx}
\usepackage{dcolumn}
\usepackage{bm}

\begin{document}

\title{Exact analytical solution of the problem of current-carrying states
of the Josephson junction in external magnetic fields}
\author{S. V. Kuplevakhsky}
\email{kuplevakhsky@ilt.kharkov.ua}
\author{A. M. Glukhov}
\affiliation{B. I. Verkin Institute for Low Temperature Physics and Engineering, \\
National Academy of Sciences of Ukraine, \\
47 Lenin Ave., 61103 Kharkov, UKRAINE}
\date{\today }

\begin{abstract}
The classical problem of the Josephson junction of arbitrary length $W$ in
the presence of externally applied magnetic fields ($H$) and transport
currents ($J$) is reconsidered from the point of view of stability theory.
In particular, we derive the complete infinite set of exact analytical
solutions for the phase difference that describe the current-carrying states
of the junction with arbitrary $W$ and an arbitrary mode of the injection of
$J$. These solutions are parameterized by two natural parameters: the
constants of integration. The boundaries of their stability regions in the
parametric plane are determined by a corresponding infinite set of exact
functional equations. Being mapped to the physical plane $\left( H,J\right) $%
, these boundaries yield the dependence of the critical transport current $%
J_{c}$ on $H$. Contrary to a wide-spread belief, the \textit{exact}
analytical dependence $J_{c}=J_{c}\left( H\right) $ proves to be multivalued
even for arbitrarily small $W$. What is more, the exact solution reveals the
existence of \textit{unquantized} Josephson vortices carrying fractional
flux and located near one of the junction edges, provided that $J$ is
sufficiently close to $J_{c}$ for certain finite values of $H$. This
conclusion (as well as other exact analytical results) is illustrated by a
graphical analysis of typical cases.
\end{abstract}

\pacs{74.50.+r, 03.75.Lm, 02.30.Oz} \maketitle

\section{Introduction}

Based on mathematical methods of stability theory, we reconsider the
classical physical problem\cite{S72,BP82,L86} of current-carrying states of
the Josephson junction of arbitrary length $W$ in external magnetic fields.
Although the problem was first posed over four decades ago\cite%
{J65,ISS66,OS67} and ever since has found numerous practical applications,%
\cite{S72,BP82,L86,r1} its complete analytical solution has not been
obtained in the previous literature. Here, we derive this solution and show
that it leads to new and important physical conclusions: the multivaluedness
of the \textit{exact} analytical dependence of the critical transport
current on the applied field for arbitrarily small $W$, and the existence of
\textit{unquantized} Josephson vortices carrying fractional flux. This paper
can be considered as a logical continuation of the investigation initiated
in our preceding publication,\cite{KG06} where we have derived the complete
analytical solution for the Josephson junction in external magnetic fields
in the absence of transport currents.

\begin{figure}
\includegraphics{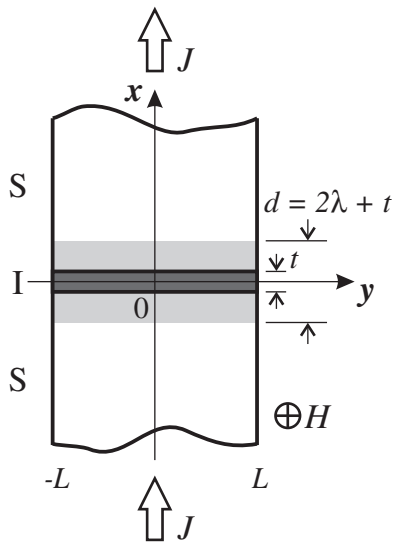}
\caption{\label{fig:junction}The geometry of the problem: $t$ is the thickness of the barrier; $%
W=2L $ is the length of the barrier; $\lambda $ is the London
penetration depth; $d=2\lambda +t$ is the width of the
field-penetration region (shaded). The external magnetic field $H$
is directed into the plane of the figure, and the transport
current $J$ is along the axis $x$.}
\end{figure}

To remind the reader of the standard formulation of the problem,
we consider the geometry presented in Fig.~\ref{fig:junction}.
Here, the $x$ axis is perpendicular to the
insulating layer $I$ (the barrier) between two identical superconductors $S$%
; the $y$ axis is along the barrier whose length is $W=2L\in \left( 0,\infty
\right) $. A constant, homogeneous external magnetic field $\mathbf{H}$ is
applied along the axis $z$: $\mathbf{H}=\left( 0,0,H\geq 0\right) $. Full
homogeneity along the $z$ axis is assumed. The transport current $\mathbf{J}$
is injected along the axis $x$: $\mathbf{J}=\left( J,0,0\right).$

In the region of field penetration, the electrodynamics of the junction in
equilibrium is fully described by a time-independent phase difference at the
barrier, $\phi =\phi \left( y\right) $. Using the dimensionless units
introduced in Ref. \cite{KG06}, we can write down the local magnetic field
and the Josephson current density as\cite{J65}%
\begin{equation}
h\left( y\right) =\frac{1}{2}\frac{d\phi }{dy}  \label{0/1}
\end{equation}%
and%
\begin{equation}
j\left( y\right) =\frac{1}{2}\sin \phi \text{,}  \label{0/2}
\end{equation}%
respectively. Accordingly, the equation for the phase difference (the
Maxwell equation) reads:%
\begin{equation}
\frac{d^{2}\phi }{dy^{2}}=\sin \phi .  \label{1/3}
\end{equation}%
Boundary conditions to (\ref{1/3}) depend on the mode of the injection of
the transport current
\[
J=\int_{-L}^{L}dyj\left( y\right) .
\]%
If it is symmetric with respect to the plane $\left( y,z\right) $, we have:
\begin{equation}
\frac{d\phi }{dy}\left( \pm L\right) =2H\pm J,  \label{1/4}
\end{equation}%
or, equivalently,\cite{ISS66,OS67}%
\begin{eqnarray}
H=\frac{1}{4}\left[ \frac{d\phi }{dy}\left( +L\right) +\frac{d\phi }{dy}%
\left( -L\right) \right] ,  \label{1/5}
\\
J=\frac{1}{2}\left[ \frac{d\phi }{dy}\left( +L\right) -\frac{d\phi }{dy}%
\left( -L\right) \right] .  \label{1/6}
\end{eqnarray}%
Solutions to (\ref{1/3}), (\ref{1/4}) are supposed to satisfy an
obvious physical requirement: they must be stable with respect to
any infinitesimal perturbations. (Unstable solutions that do not
meet this requirement are physically unobservable and should be
rejected.)

Unfortunately, the standard boundary-value problem (\ref{1/3}), (\ref{1/4})
is mathematically ill-posed:\cite{CH} (i) for $\left\vert J\right\vert $
larger than certain $J_{\max }=J_{\max }\left( H,L\right) $, it does not
admit any solutions at all; (ii) aside from stable (physical) solutions,
there may exist unstable (unphysical) solutions for the same $H$ and $J$;
(iii) for the same $H$ and $J$, there may exist several different physical
solutions. An immediate consequence of this ill-posedness is as follows:
although the general solution to (\ref{1/3}) is well-known,\cite{A70} the
constants of integration specifying particular physical solutions cannot be
determined directly from the boundary conditions (\ref{1/4}).

In view of the above-mentioned mathematical difficulties, the previous
analysis of the problem (\ref{1/3}), (\ref{1/4}) was concentrated mainly on
finding the dependence $J_{\max }=J_{\max }\left( H\right) $ (for particular
values of $L$) without trying to establish the exact analytical form of
current-carrying solutions. (It should be noted that the quantity $J_{\max }$
itself was identified with the experimentally observable critical current $%
J_{c}$, i.e., the identity $J_{\max }\equiv J_{c}$ was assumed.)

For the case $L\ll 1$, there existed\cite{J65} a simple analytical
approximation for the dependence $J_{\max }=J_{\max }\left( H\right) $ (the
so-called\cite{S72,BP82,L86} "Fraunhofer pattern"). As to the case $L\gtrsim
1$, only particular numerical results were obtained. Thus, Owen and Scalapino%
\cite{OS67} established the dependence $J_{\max }=J_{\max }\left( H\right) $
only for $L=5$: it proved to be multivalued. The numerical method of Ref.
\cite{OS67} was later employed to study the effect of asymmetric injection
of the transport current.\cite{BB75} Unfortunately, all these numerical
results could tell very little about general properties of the
current-carrying states for arbitrary $L\in \left( 0,\infty \right) $.
Besides, no analytical expressions were derived that could serve for direct
determination of $J_{\max }$.

On the other hand, attempts were made\cite{Zh78,ZhZ78} to simplify the
computational procedure\cite{OS67} by transforming the boundary-value
problem (\ref{1/3}), (\ref{1/4}) into an equivalent initial-value problem.
Although these attempts did not produce exact analytical solutions, we note
that Refs. \cite{Zh78,ZhZ78} introduced a new, more satisfactory
mathematical definition of the observable critical current $J_{c}$: it was
identified with the boundary of the stability regions of the
current-carrying configurations. The same mathematical definition of $J_{c}$
was employed in Refs. \cite{Ga84,Se04} concerned with certain nontrivial
generalizations of the boundary-value problem (\ref{1/3}), (\ref{1/4}).
Unfortunately, exact analytical expressions for the physical solutions to (%
\ref{1/3}), (\ref{1/4}) were not found in Refs. \cite{Ga84,Se04}, either.

As already mentioned, in Ref. \cite{KG06} we have derived the complete
infinite set of exact physical solutions to (\ref{1/3}), (\ref{1/4}) under
the condition $J=0$. The approach of Ref. \cite{KG06} consists in a certain
generalization of the boundary conditions and an application of methods of
stability theory at an early stage of the consideration. The same approach
is adopted in this paper for the general case $J\neq 0$. Thus, we derive a
complete set of exact particular solutions to (\ref{1/3}) that are stable
under the condition that $\frac{d\phi }{dy}$ is fixed at the boundaries $%
y=\pm L$ [for arbitrary $L\in \left( 0,\infty \right) $]. These solutions
are parameterized by two natural parameters: the constants of integration of
(\ref{1/3}). The boundaries of their stability regions are determined by a
corresponding infinite set of exact functional equations. The physical
interpretation of the obtained solutions stems from the fact that the
boundary conditions in the form (\ref{1/5}), (\ref{1/6}) (or their
modification for the case of asymmetric injection of $J$) realize a mapping
of the stability regions from the parametric plane to the physical plane $%
\left( H,J\right) $.

In Sec. II, we present a static method of the analysis of stability based on
the minimization of the generating free-energy functional. A Sturm-Liouville
eigenvalue problem that plays a key role in the analysis of stability is
discussed. In Sec. III, we derive the complete set of exact stable
analytical solutions to (\ref{1/3}), (\ref{1/4}) under the condition $H\geq
0 $, $J\geq 0$. A numerical analysis of several typical cases is carried
out. In Sec. IV, we elaborate on major physical implications of the exact
analytical solutions. Graphic illustrations are presented. Generalizations
to the case of arbitrary sign of $H$ and $J$, and to the case of asymmetric
injection of $J$ are considered. Finally, in Sec. V, we summarize the
obtained physical and mathematical results and make several concluding
remarks.

In Appendix A, an alternative (dynamic) method of the analysis of stability
is presented. In Appendix B, functional equations for the stability regions
are derived. In Appendix C, a certain special solution of the
Sturm-Liouville eigenvalue problem is considered.

\section{Analysis of stability}

The stability of the solutions to (\ref{1/3})-(\ref{1/6}) can be analyzed by
means of two different methods: a static\cite{KG06} one, and a dynamic\cite%
{JJ80} one. Although they are fully equivalent mathematically, the static
method seems to be more natural physically: we therefore discuss it in this
section. (For the sake of completeness, we outline the dynamic method in
Appendix A.)

\subsection{Minimization of the Gibbs free-energy functional}

The static method is based on the minimization of the generating Gibbs
free-energy functional. For the boundary-value problem (\ref{1/3}), (\ref%
{1/4}), the corresponding functional (in terms of the dimensionless units,%
\cite{KG06} and per unit length along the $z$ axis) has the following form:
\begin{equation}
\Omega _{G}\left[ \phi ,\frac{d\phi }{dy};H,J\right] =2H^{2}W+\int_{-L}^{L}dy%
\left[ 1-\cos \phi \left( y\right) +\frac{1}{2}\left[ \frac{d\phi
\left( y\right) }{dy}\right] ^{2}\right] -\left( 2H+J\right) \phi
\left( L\right) +\left( 2H-J\right) \phi \left( -L\right) .
\label{1/1}
\end{equation}%
As can be easily seen, the stationarity condition of (\ref{1/1}),%
\[
\delta \Omega _{G}\left[ \phi ,\frac{d\phi }{dy};H,J\right] =0,
\]%
yields the equation for the phase difference (\ref{1/3}) and the boundary
conditions (\ref{1/4}).

Note that the functional (\ref{1/1}) with $J=0$ is analyzed in Ref. \cite%
{KG06}. Basic properties of functionals of the type (\ref{1/1}) are also
discussed in Refs. \cite{K04,K05}: in particular, all the stationary points
of (\ref{1/1}) are either local minima or saddle points.\cite{r2}

In full analogy with the case $J=0$,\cite{KG06} the type of a stationary
point $\phi =\phi \left( y\right) $ obeying (\ref{1/3}), (\ref{1/4}) is
determined by the sign of the lowest eigenvalue $\mu =\mu _{0}$ of the
Sturm-Liouville problem%
\begin{eqnarray}
-\frac{d^{2}\psi }{dy^{2}}+\cos \phi \left( y\right) \psi =\mu \psi ,\quad
y\in \left( -L,L\right) ,  \label{1/7}
\\
\frac{d\psi }{dy}\left( -L\right) =\frac{d\psi }{dy}\left( L\right) =0,
\label{1/8}
\end{eqnarray}%
Namely, if $\mu _{0}<0$, the solution $\phi =\phi \left( y\right)
$ corresponds to a saddle point of (\ref{1/1}) ($\delta \Omega
_{G}^{2}\gtrless 0$). Solutions of this type are absolutely
unstable and hence unphysical.

On the contrary, the stable physical solutions $\phi =\phi \left( y\right) $
that minimize (\ref{1/1}) are characterized by $\mu _{0}>0$ ($\delta \Omega
_{G}^{2}>0$). The boundaries of the stability regions for these solutions ($%
\delta \Omega _{G}^{2}\geq 0$) are determined by the condition%
\[
\mu _{0}=0,
\]%
or, equivalently, by the solution $\bar{\psi}_{0}=\bar{\psi}_{0}\left(
y\right) $ to the boundary-value problem%
\begin{eqnarray}
-\frac{d^{2}\bar{\psi}_{0}}{dy^{2}}+\cos \phi \left( y\right) \bar{\psi}%
_{0}=0,\quad y\in \left( -L,L\right) ,  \label{1/11}
\\
\frac{d\bar{\psi}_{0}}{dy}\left( -L\right) =\frac{d\bar{\psi}_{0}}{dy}\left(
L\right) =0,  \label{1/12}
\\
\bar{\psi}_{0}\left( y\right) \neq 0,\quad y\in \left[ -L,L\right] .
\label{1/13}
\end{eqnarray}

Equation (\ref{1/7}) can be transformed into Lam\'{e}'s equation.\cite{WW27}
In certain limiting cases, the eigenvalue $\mu =\mu _{0}$ (and the
corresponding eigenfunction $\psi =\psi _{0}$) of the problem (\ref{1/7}), (%
\ref{1/8}) can be found explicitly by perturbation methods: see Appendix C.
However, since we will mostly need information about the boundaries of the
stability regions, the consideration of the main part of this paper is based
on the fact that the linear boundary-value problem (\ref{1/11})-(\ref{1/13})
is exactly solvable. The relevant exact analytical solutions are derived in
Appendix B.

\section{Current-carrying states}

As is well-known,\cite{A70} the general solution to (\ref{1/3}) can be
easily obtained using the first integral,
\begin{equation}
\frac{1}{2}\left[ \frac{d\phi }{dy}\right] ^{2}+\cos \phi =C,\quad -1\leq
C<\infty ,  \label{2/1}
\end{equation}%
where $C\,$ is the constant of integration. In Ref. \cite{KG06}, we have
written down the general solution to (\ref{1/3}) in the form convenient for
applications with the boundary conditions (\ref{1/4}). In that paper,
solutions parameterized by $C\in \left[ -1,1\right) $ and $C\in \left(
1,+\infty \right) $ have been termed solutions of type I and type II,
respectively.

As we have shown for $H\neq 0$, $J=0$,\cite{KG06} all the solutions of type
I are absolutely unstable. On the contrary, the solutions of type II
contain, for $H\neq 0$, $J=0$, a subclass of stable solutions.

The case $H\neq 0$, $J\neq 0$ is quite different, because both the
classes of solutions (of type I and type II) contain subclasses of
stable current-carrying solutions. [For example, for $J=2H$, we
have $C=\cos \phi \left( -L\right) <1$, since $\phi \left(
-L\right) \neq 0 \text{ mod } 2\pi $.]
In view of continuous dependence of the left-hand side of (\ref{2/1}) on $C$%
, stable current-carrying solutions of type I in the limit $C\rightarrow 1-0$
should coincide with stable current-carrying solutions of type II obtained
by the limiting procedure $C\rightarrow 1+0$.

Note that, in what follows, we will employ instead of $C$ a standard
parametrization constant $k$.\cite{KG06} Namely,%
\begin{equation}
k^{2}\equiv \frac{1+C}{2},\quad 0\leq k<1  \label{2/2}
\end{equation}%
for the solutions\ of type I, and%
\begin{equation}
k^{2}\equiv \frac{2}{1+C},\quad 0<k<1  \label{2/3}
\end{equation}%
for the solutions\ of type II. Moreover, in this section, we
restrict ourselves to symmetric injection of $J$ [conditions
(\ref{1/4})], and to the case $H\geq 0$, $J\geq 0$. (These
restrictions will be removed in Sec. IV.)

According to the scheme outlined in the Introduction, we start with finding
all the solutions to (\ref{1/3}) that are stable under the condition that $%
\frac{d\phi }{dy}$ is fixed at the boundaries $y=\pm L$. These
solutions are parameterized by $k$ and the second (additive)
constant of integration denoted as $\beta $ (for solution of type
I) or $\alpha $ (for solutions of type II). The boundaries of the
stability regions are determined from the solution to the linear
boundary-value problem (\ref{1/11})-(\ref{1/13}). Finally,
relations (\ref{1/5}), (\ref{1/6}) are employed to map the
stability regions from the parametric planes $\left( k,\beta \right) $ and $%
\left( k,\alpha \right) $ to the physical plane $\left( H,J\right) $.

\subsection{Solutions of type I}

The general form of the solutions of type I is given by\cite{KG06}%
\begin{equation}
\phi _{\pm }\left( y\right) =\pi \left( 2n+1\right) \pm 2\arcsin \left[ k\,%
\text{sn}\left( y-y_{0},k\right) \right] ,\quad n=0,\pm 1,\ldots ,
\label{2/4}
\end{equation}%
where $\text{sn }u$ is the Jacobian elliptic sine.\cite{AS65} The
constant of
integration $y_{0}$ is subject to the restriction%
\begin{equation}
-K\left( k\right) \leq y_{0}<K\left( k\right) ,  \label{2/5}
\end{equation}%
with $K\left( k\right) $ being the complete elliptic integral of the first
kind,\cite{AS65} and the constant of integration $k$ is defined by (\ref{2/2}%
). Taking into account that $k=0$ in (\ref{2/4}) corresponds to absolutely
unstable solutions with $H=J=0$,\cite{KG06} we impose the condition%
\begin{equation}
0<k<1.  \label{2/6}
\end{equation}

Mathematically, it is convenient to begin the consideration of the
current-carrying solutions of type I with the case $H=0$, $J\geq
0$. The solutions for the case $H\geq 0$, $J\geq 0$ will be
obtained from the solutions for $H=0$, $J\geq 0$ by the
introduction of a new parameter.

\subsubsection{The case $H=0$, $J\geq 0$}

The generalized form of the boundary conditions (\ref{1/4}) for $H=0$, $J>0$
is given by the relations%
\begin{eqnarray}
\frac{d\phi }{dy}\left( L\right) =-\frac{d\phi }{dy}\left( -L\right) ,
\label{2/7}
\\
\frac{d\phi }{dy}\left( L\right) =\text{const}>0.  \label{2/8}
\end{eqnarray}%
Using (\ref{2/7}), we find that $y_{0}=-K\left( k\right) $ in
(\ref{2/4}), whereas (\ref{2/8}) yields $\phi \equiv \phi _{-}$
[$L<2K\left( k\right) $].
Finally, setting $n=0$ in (\ref{2/4}), we obtain%
\begin{equation}
\phi _{s}\left( y\right) =2\arccos \left[ k\,\frac{\text{cn}\left(
y,k\right) }{\text{dn}\left( y,k\right) }\right] ,  \label{2/9}
\end{equation}%
where $\text{cn }u$ and $\text{dn }u$ are the Jacobian elliptic
cosine and the delta amplitude, respectively.\cite{AS65}

This solution is symmetric with respect to reflection:%
\begin{equation}
\phi _{s}\left( -y\right) =\phi _{s}\left( y\right) .  \label{2/9.1}
\end{equation}%
It is stable only for%
\begin{equation}
k\in \left[ k_{c},1\right] ,  \label{2/10}
\end{equation}%
where, according to the results of Appendix B, the boundary of the stability
region $k_{c}=k_{c}\left( L\right) $ is implicitly determined by the
functional equation%
\begin{equation}
\text{cn}\left( L,k_{c}\right) \left[ -E\left( L,k_{c}\right)
+\left( 1-k_{c}^{2}\right) L\right] +\left( 1-k_{c}^{2}\right)
\text{sn}\left( L,k_{c}\right) \text{dn}\left( L,k_{c}\right) =0,
\label{2/11}
\end{equation}%
with $E\left( u,k\right) $ being the incomplete elliptic integral of the
first kind.\cite{AS65} [We include the point $k=1$ in the definition of the
stability region (\ref{2/10}), because lim$_{k\rightarrow 1}\phi _{s}\equiv
0 $, which is an absolutely stable solution for the case $H=J=0$.]

\begin{figure}
\includegraphics{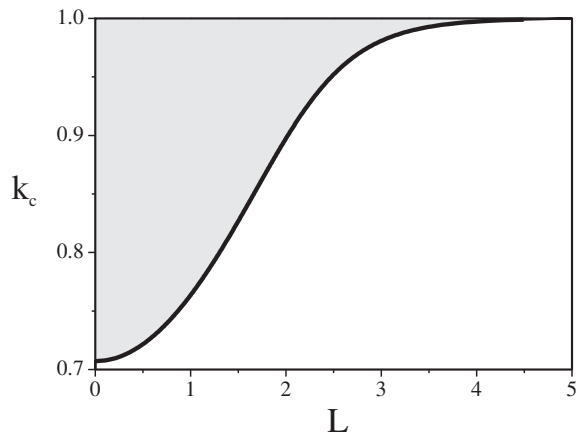}
\caption{\label{fig:kc(L)}The dependence $k_{c}=k_{c}\left(
L\right) $ (solid line). The stability region is shaded.}
\end{figure}

Equation (\ref{2/11}) can be solved analytically in two limiting cases. In
particular, for $L\ll 1$, the solution is%
\begin{equation}
k_{c}\thickapprox \frac{1}{\sqrt{2}}.  \label{2/12}
\end{equation}%
For $L\gg 1$, equation (\ref{2/11}) becomes%
\begin{equation}
K\left( k_{c}\right) \thickapprox L,  \label{2/13}
\end{equation}%
and the solution is%
\begin{equation}
k_{c}\thickapprox 1-8\exp \left( -2L\right) .  \label{2/14}
\end{equation}%
For arbitrary $L\in \left( 0,\infty \right) $, we present the
numerical solution to (\ref{2/11}) in Fig.~\ref{fig:kc(L)}.

Substituting (\ref{2/9}) into (\ref{1/6}), we arrive at the
expression for
the current $J=J\left( L,k\right) $:%
\begin{equation}
J=2k\sqrt{1-k^{2}}\frac{\text{sn}\left( L,k\right)
}{\text{dn}\left( L,k\right) },\quad k\in \left[ k_{c},1\right] .
\label{2/15}
\end{equation}%
Note that for $L\equiv \frac{W}{2}\ll 1$ expression (\ref{2/15}) reduces to
the expected result\cite{J65,S72,BP82,L86}%
\begin{equation}
J\thickapprox \frac{W}{2}\sin \phi _{s}\left( 0\right) ,  \label{2/16}
\end{equation}%
where, by (\ref{2/9}) and (\ref{2/12}),%
\[
\phi _{s}\left( 0\right) =2\arccos k\in \left[ 0,\frac{\pi }{2}\right] .
\]

\begin{figure}
\includegraphics{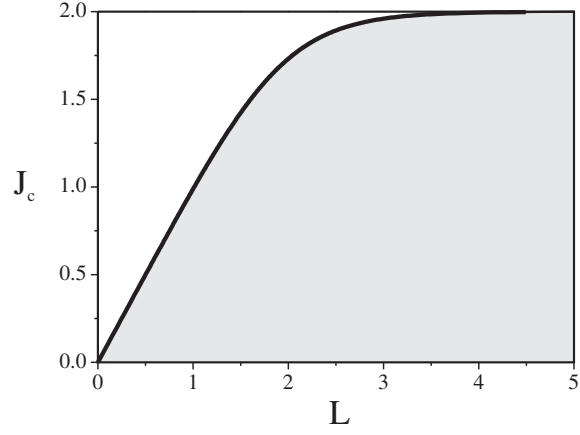}
\caption{\label{fig:Jc(L)}The dependence $J_{c}=J_{c}\left(
L\right) $ for $H=0$ (solid line). The stability region is
shaded.}
\end{figure}

According to (\ref{2/15}), the dependence $J_{c}=J_{c}\left( L\right) $ is
given by%
\begin{equation}
J_{c}=2k_{c}\sqrt{1-k_{c}^{2}}\frac{\text{sn}\left( L,k_{c}\right) }{\text{dn%
}\left( L,k_{c}\right) }.  \label{2/17}
\end{equation}%
Thus, for $L\equiv \frac{W}{2}\gg 1$, we get, using (\ref{2/13}), (\ref{2/14}%
),%
\begin{equation}
J_{c}\thickapprox 2\left[ 1-8\exp \left( -W\right) \right] .  \label{2/18}
\end{equation}%
For arbitrary $L\in \left( 0,\infty \right) $, the dependence $%
J_{c}=J_{c}\left( L\right) $ is presented in Fig.~\ref{fig:Jc(L)}.
Although Fig.~\ref{fig:Jc(L)}
reproduces the old results\cite{OS67} obtained by numerical maximization of $%
J$, we want to emphasize a substantial methodological difference: the curve $%
J_{c}=J_{c}\left( L\right) $ in Fig.~\ref{fig:Jc(L)} is nothing but a mapping by means of (%
\ref{2/17}) of the boundary of the stability region
$k_{c}=k_{c}\left( L\right) $ in Fig.~\ref{fig:kc(L)}.

\subsubsection{The case $H\geq 0$, $J\geq 0$}

For $H>0$, $J>0$, instead of (\ref{2/7}), we have%
\begin{equation}
\frac{d\phi }{dy}\left( -L\right) =\text{const},\quad \frac{d\phi }{dy}%
\left( -L\right) \neq \pm \frac{d\phi }{dy}\left( L\right) .  \label{2/19}
\end{equation}%
Boundary conditions (\ref{2/19}) break the symmetry (\ref{2/9.1}). Taking
into account that in the limit $H\rightarrow 0$ we must get (\ref{2/9}),
conditions (\ref{2/8}) and (\ref{2/19}) can be satisfied by%
\begin{equation}
\phi _{s}\left( y\right) =2\arccos \left[ k\,\frac{\text{cn}\left(
y+\beta ,k\right) }{\text{dn}\left( y+\beta ,k\right)
}\right],\quad k\in \left[ k_{c},1\right) ,\quad \beta \in \left[
0,\beta _{c}\right] ,
\label{2/20}
\end{equation}%
where $k_{c}$ is determined by (\ref{2/11}), and $\beta _{c}\in
\left[ 0,K\left( k\right) \right) $. The boundary of the stability
region $\beta _{c}=\beta _{c}\left( k\right) $ is determined (see
Appendix B) by the
solution to the functional equation%
\begin{eqnarray}
\text{cn}\left( L+\beta _{c},k\right) \text{cn}\left( L-\beta _{c},k\right) %
\left[ -E\left( L+\beta _{c},k\right) -E\left( L-\beta
_{c},k\right)+\left( 1-k^{2}\right) L\right]  \nonumber \\
+\left( 1-k^{2}\right) \left[ \text{sn}\left( L+\beta _{c},k\right) \text{cn}%
\left( L-\beta _{c},k\right) \text{dn}\left( L+\beta _{c},k\right)\right.  \nonumber \\
\left.+\text{sn}\left( L-\beta _{c},k\right) \text{cn}\left(
L+\beta_{c},k\right) \text{dn}\left( L-\beta _{c},k\right) \right] =0,\quad k\in %
\left[ k_{c},1\right) ,  \label{2/21}
\end{eqnarray}%
under the condition $\beta _{c}\left( k_{c}\right) =0$.

\begin{figure}
\includegraphics{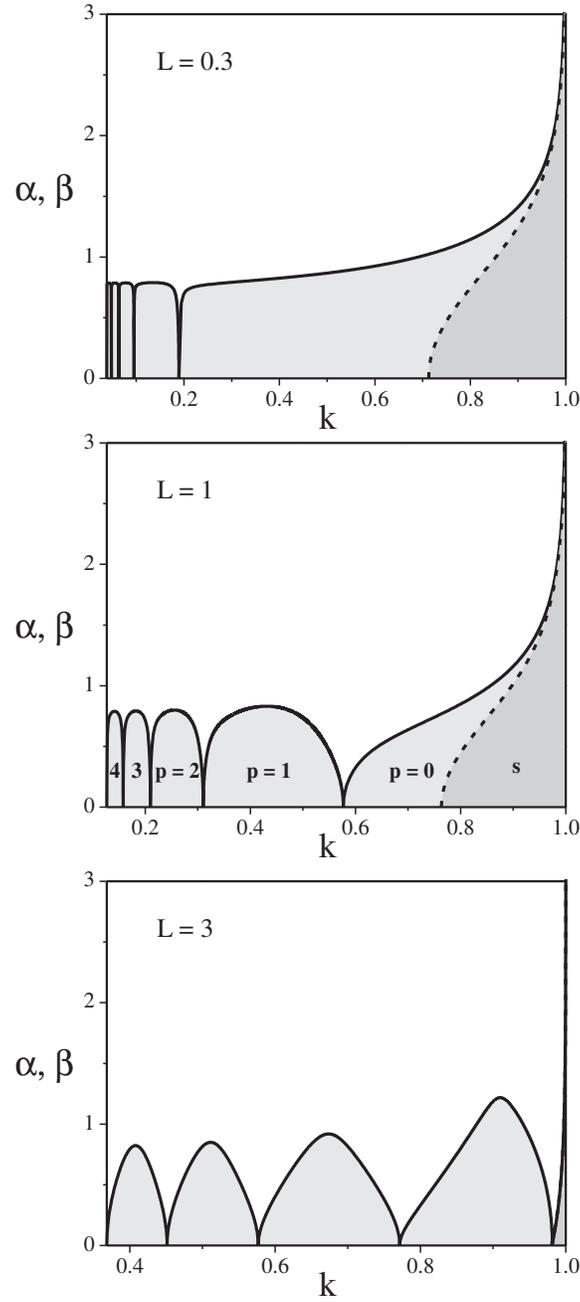}
\caption{\label{fig:a&b(k)}The stability regions of $\phi _{s}$ and $\phi _{p}$ ($p=0,1,2\ldots $%
) in the parametric plane (shaded) for $L=0.3,1,3$. The
dependencies $\beta _{c}=\beta _{c}\left( k\right) $ and $\alpha
_{c}=\alpha _{c}\left( k\right) $ are given by the dashed line and
the solid lines, respectively.}
\end{figure}

In Fig.~\ref{fig:a&b(k)}, we present the stability region of
(\ref{2/20}) obtained by
numerical evaluation of Eq. (\ref{2/21}) for several different values of $L$%
: $L=0.3$ (a "small" junction), $L=1$ (a "medium" junction), and $L=3$ (a
"large" junction). As we can see, $\lim_{k\rightarrow 1}\beta _{c}\left(
k\right) \rightarrow \infty $. The asymptotics of $\beta _{c}\left( k\right)
$ for $k\rightarrow 1$ can be established analytically.

Let us make the substitution%
\begin{equation}
\beta _{c}=K\left( k\right) -\gamma _{c}  \label{2/21.1}
\end{equation}%
in Eq. (\ref{2/21}). By proceeding to the limit $k=1$, we obtain a
functional equation that determines the dependence $\gamma _{c}=\gamma
_{c}\left( L\right) $ for $k=1$:%
\begin{eqnarray}
L\sinh \left( L-\gamma _{c}\right) \sinh \left( L+\gamma
_{c}\right) -\frac{1}{2}\sinh ^{2}\left( L-\gamma _{c}\right)
\sinh \left( L+\gamma _{c}\right) \cosh \left( L-\gamma
_{c}\right)  \nonumber \\
-\frac{1}{2}\sinh ^{2}\left( L+\gamma _{c}\right) \sinh \left(
L-\gamma _{c}\right) \cosh \left( L+\gamma _{c}\right) -\sinh
\left( L+\gamma _{c}\right) \cosh \left( L-\gamma _{c}\right)
-\sinh \left( L-\gamma _{c}\right) \cosh \left( L+\gamma
_{c}\right) =0. \label{2/21.2}
\end{eqnarray}%
\begin{figure}
\includegraphics{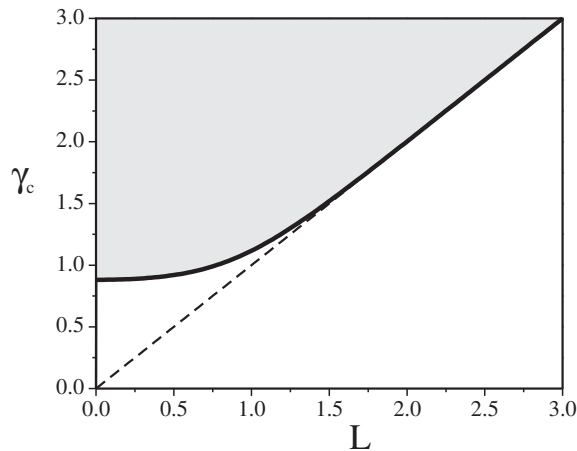}
\caption{\label{fig:gc(L)}The dependence $\gamma _{c}=\gamma
_{c}\left( L\right) $ (solid line). The stability region is
shaded.}
\end{figure}
The numerical solution to this equation is given in Fig.~\ref{fig:gc(L)}. [Note that $%
\gamma _{c}\left( L\right) \thickapprox L$ for $L\gg 1$.] Taking
into
account relation (\ref{2/21.1}), we arrive at the sought asymptotics of $%
\beta _{c}\left( k\right) $ for $k\rightarrow 1$:%
\begin{equation}
\beta _{c}\left( k\right) \thickapprox \frac{1}{2}\ln \frac{16}{1-k^{2}}%
-\gamma _{c}\left( L\right) .  \label{2/21.3}
\end{equation}%
Accordingly, the limiting form of the current-carrying solution (\ref{2/20})
is%
\begin{equation}
\lim_{k\rightarrow 1}\phi _{s}\left( y\right) \equiv \phi _{l}\left(
y\right) =4\arctan \left[ \exp \left( y-\gamma \right) \right] ,
\label{2/21.4}
\end{equation}%
where $\gamma \in \left[ \gamma _{c},\infty \right) $ (see
Fig.~\ref{fig:gc(L)}).

Equations (\ref{1/5}) and (\ref{1/6}), upon the substitution of (\ref{2/20}%
), yield%
\begin{eqnarray}
H=\frac{k}{2}\sqrt{1-k^{2}}\left[ \frac{\text{sn}\left( L+\beta ,k\right) }{%
\text{dn}\left( L+\beta ,k\right) }-\frac{\text{sn}\left( L-\beta
,k\right) }{\text{dn}\left( L-\beta ,k\right) }\right] ,
\label{2/22}
\\
J=k\sqrt{1-k^{2}}\left[ \frac{\text{sn}\left( L+\beta ,k\right) }{\text{dn}%
\left( L+\beta ,k\right) }+\frac{\text{sn}\left( L-\beta ,k\right) }{\text{dn%
}\left( L-\beta ,k\right) }\right] , \label{2/23}
\\
k\in\left[k_{c},1\right) , \quad \beta \in
\left[0,\beta_{c}\right] . \nonumber
\end{eqnarray}%
In the limit $k=1$, these relations take the form%
\begin{eqnarray}
H=\frac{\cosh L\cosh \gamma }{\cosh \left( \gamma -L\right) \cosh
\gamma }, \label{2/24}
\\
J=\frac{2\sinh L\sinh \gamma }{\cosh \left( \gamma -L\right) \cosh \gamma }%
,\quad \gamma \in \left[ \gamma _{c},\infty \right) .  \label{2/25}
\end{eqnarray}%
By setting $\beta =\beta _{c}$ in (\ref{2/22}) and (\ref{2/23}),
we can obtain a relevant part of the dependence $J_{c}=J_{c}\left(
H\right) $ for arbitrary $L\in \left( 0,\infty \right) $: see Sec.
IV.

We conclude the discussion of the solution $\phi _{s}$ by presenting an
explicit analytical expression for the special case $J=2H$ that was the
subject of numerical evaluation in Ref. \cite{ZhZ78}. From (\ref{2/22}) and (%
\ref{2/23}), we find: $\beta =L$. Substitution into (\ref{2/20}) immediately
yields%
\begin{equation}
\left. \phi _{s}\left( y\right) \right\vert _{J=2H}=2\arccos \left[ k\,\frac{%
\text{cn}\left( y+L,k\right) }{\text{dn}\left( y+L,k\right)
}\right] ,\quad k\in \left[ k_{m},1\right] ,  \label{2/25.1}
\end{equation}%
where $k_{m}$ is determined by the condition $\beta _{c}\left( k_{m}\right)
=L$.

\subsection{Solutions of type II}

We start with the stable type-II solutions for the case $H\geq 0$, $J=0$:%
\cite{K04,K05,KG06}
\begin{eqnarray}
\phi _{p}(y)=\pi \left( p-1\right) +2\text{am}\left(
\frac{y}{k}+K\left( k\right) ,k\right) , \quad p=2m \quad \left(
m=0,1,\ldots \right) ; \label{2/26}
\\
\phi _{p}(y)=\pi p+2\text{am}\left( \frac{y}{k},k\right) , \quad
p=2m+1 \quad \left( m=0,1,\ldots \right) , \label{2/27}
\end{eqnarray}%
where $\text{am }u$ is the Jacobian amplitude.\cite{AS65} The
stability
regions of (\ref{2/26}), (\ref{2/27}) are given by%
\begin{equation}
p=0:k\in \left( k_{1},1\right) ;\quad p=1,2,\ldots :k\in \left(
k_{p+1},k_{p} \right] ,  \label{2/28}
\end{equation}%
where the points $k=k_{p}$ ($p=1,2,\ldots $) are the roots of the equations%
\begin{equation}
pk_{p}K\left( k_{p}\right) =L,\quad p=1,2,\ldots .  \label{2/29}
\end{equation}%
Solutions (\ref{2/26}), (\ref{2/27}) form an infinite set, and the union of
their stability regions (\ref{2/28}) (they interchange for even and odd $p$)
is equal to the whole $k$-interval $\left( 0,1\right) $. The meaning of the
parameter $p=0,1,2,\ldots $ is revealed by the relation%
\begin{equation}
p=\left\lfloor \frac{\phi _{p}\left( L\right) -\phi _{p}\left( -L\right) }{%
2\pi }\right\rfloor ,  \label{2/29.1}
\end{equation}%
where $\left\lfloor \ldots \right\rfloor $ stands for the integer part of
the argument.\cite{r3} Note also the symmetry property:%
\begin{equation}
\phi _{p}\left( -y\right) =2\pi p-\phi _{p}\left( y\right) .  \label{2/29.2}
\end{equation}

For $H>0$, $J>0$, current-carrying type-II solutions obey the generalized
boundary conditions (\ref{2/8}), (\ref{2/19}) that break the symmetry (\ref%
{2/29.2}). These conditions can be satisfied if, in (\ref{2/26}) and (\ref%
{2/27}), we make a shift of the argument $y\rightarrow y+k\alpha $ (with $%
\alpha $ being a new parameter):
\begin{eqnarray}
\phi _{p}(y)=\pi \left( p-1\right) +2\text{am}\left(
\frac{y}{k}+K\left( k\right) +\alpha ,k\right) ,\quad \alpha \in
\left[ 0,\alpha _{c}\right] , \quad p=2m \quad \left( m=0,1,\ldots
\right) ; \label{2/30}
\\
\phi _{p}(y)=\pi p+2\text{am}\left( \frac{y}{k}+\alpha ,k\right)
,\quad \alpha \in \left[ 0,\alpha _{c}\right] , \quad p=2m+1 \quad
\left( m=0,1,\ldots \right) . \label{2/31}
\end{eqnarray}%
The domains of the parameter $k$ in (\ref{2/30}) and (\ref{2/31})
are given by (\ref{2/28}), (\ref{2/29}), whereas $\alpha _{c}\in
\left[ 0,K\left( k\right) \right) $. The boundaries of the
stability regions $\alpha _{c}=\alpha _{c}\left( k\right) $ are
determined (see Appendix B) by the
solutions to the functional equation%
\begin{eqnarray}
k^{2}\text{sn}\left( \frac{L}{k}+\alpha _{c},k\right) \text{sn}\left( \frac{L%
}{k}-\alpha _{c},k\right) \text{cn}\left( \frac{L}{k}+\alpha
_{c},k\right) \text{cn}\left( \frac{L}{k}-\alpha _{c},k\right)
\left[ E\left( \frac{L}{k}+\alpha _{c},k\right) +E\left( \frac{L}{k}%
-\alpha _{c},k\right) \right]  \nonumber \\
+\text{sn}\left( \frac{L}{k}-\alpha _{c},k\right) \text{cn}\left( \frac{L}{k}%
-\alpha _{c},k\right) \text{dn}^{3}\left( \frac{L}{k}+\alpha
_{c},k\right)
+\text{sn}\left( \frac{L}{k}+\alpha _{c},k\right) \text{cn}\left( \frac{L}{k}%
+\alpha _{c},k\right) \text{dn}^{3}\left( \frac{L}{k}-\alpha
_{c},k\right) =0 \label{2/32}
\end{eqnarray}%
in the case (\ref{2/30}), and to the functional equation%
\begin{eqnarray}
\frac{k^{2}}{1-k^{2}}\text{sn}\left( \frac{L}{k}+\alpha _{c},k\right) \text{%
sn}\left( \frac{L}{k}-\alpha _{c},k\right) \text{cn}\left( \frac{L}{k}%
+\alpha _{c},k\right) \text{cn}\left( \frac{L}{k}-\alpha
_{c},k\right)  \nonumber \\
\times \left\{ E\left( \frac{L}{k}+\alpha _{c},k\right) +E\left( \frac{L}{k}%
-\alpha _{c},k\right) -k^{2} \left[ \frac{\text{sn}\left(
\frac{L}{k}+\alpha _{c},k\right)
\text{cn}\left( \frac{L}{k}+\alpha _{c},k\right) }{\text{dn}\left( \frac{L}{k%
}+\alpha _{c},k\right) }+\frac{\text{sn}\left( \frac{L}{k}-\alpha
_{c},k\right) \text{cn}\left( \frac{L}{k}-\alpha _{c},k\right) }{\text{dn}%
\left( \frac{L}{k}-\alpha _{c},k\right) }\right] \right\}  \nonumber \\
-\frac{\text{sn}\left( \frac{L}{k}-\alpha _{c},k\right)
\text{cn}\left( \frac{L}{k}-\alpha _{c},k\right) }{\text{dn}\left(
\frac{L}{k}+\alpha
_{c},k\right) }-\frac{\text{sn}\left( \frac{L}{k}+\alpha _{c},k\right) \text{%
cn}\left( \frac{L}{k}+\alpha _{c},k\right) }{\text{dn}\left( \frac{L}{k}%
-\alpha _{c},k\right) }=0  \label{2/33}
\end{eqnarray}%
in the case (\ref{2/31}). The relevant solutions to (\ref{2/32}) and (\ref%
{2/33}) must satisfy the conditions $\alpha _{c}\left( k_{p}\right) =0$ ($%
p=1,2,\ldots $).

Making the substitution%
\begin{equation}
\alpha _{c}=K\left( k\right) -\gamma _{c}  \label{2/33.1}
\end{equation}%
and proceeding to the limit $k=1$ in Eq. (\ref{2/32}) with $p=0$, we arrive
at Eq. (\ref{2/21.2}). Accordingly, for $k\rightarrow 1$, the asymptotics of
$\alpha _{c}\left( k\right) $ coincides with that of $\beta _{c}\left(
k\right) $ [relation (\ref{2/21.3})]:%
\begin{equation}
\alpha _{c}\left( k\right) \thickapprox \frac{1}{2}\ln \frac{16}{1-k^{2}}%
-\gamma _{c}\left( L\right) ,  \label{2/33.2}
\end{equation}%
where the dependence $\gamma _{c}=\gamma _{c}\left( L\right) $ is
represented by the graph in Fig.~\ref{fig:gc(L)}. The feature
(\ref{2/33.2}) is clearly reproduced in Fig.~\ref{fig:a&b(k)},
where we present the stability regions of (\ref{2/30})
and (\ref{2/31}) obtained by numerical evaluation of (\ref{2/32}) and (\ref%
{2/33}), respectively, for $L=0.3,1,3$. Moreover, as could by expected from
the general arguments at the beginning of this section, the limiting form ($%
k=1$) of the current-carrying solution (\ref{2/30}) for $p=0$ coincides with
the limiting form ($k=1$) of the current-carrying solution (\ref{2/20}),
i.e.,%
\[
\lim_{k\rightarrow 1}\phi _{0}=\lim_{k\rightarrow 1}\phi _{s}=\phi _{l},
\]%
where $\phi _{l}=\phi _{l}\left( y\right) $ is given by (\ref{2/21.4}).

It is interesting to note that equations (\ref{2/32}) and (\ref{2/33}) have
exact analytical solutions at the points $k=k_{p}^{\ast }$ ($p=0,1,2,\ldots $%
), where $k_{p}^{\ast }$ are implicitly determined by the equations%
\begin{equation}
\left( p+\frac{1}{2}\right) k_{p}^{\ast }K\left( k_{p}^{\ast }\right)
=L,\quad p=0,1,2,\ldots .  \label{2/33.3}
\end{equation}%
Namely,%
\begin{equation}
\alpha _{c}\left( k_{p}^{\ast }\right) =\frac{1}{2}K\left( k_{p}^{\ast
}\right) ,\quad p=0,1,2,\ldots .  \label{2/33.4}
\end{equation}%
The role of these solutions is discussed in Sec. IV.

Upon the substitution of (\ref{2/30}) and (\ref{2/31}) into (\ref{1/5}) and (%
\ref{1/6}), we obtain, respectively,%
\begin{eqnarray}
H=\frac{\sqrt{1-k^{2}}}{2k}\left[ \text{dn}^{-1}\left(
\frac{L}{k}+\alpha ,k\right) +\text{dn}^{-1}\left(
\frac{L}{k}-\alpha ,k\right) \right] , \label{2/34}
\\
J=\frac{\sqrt{1-k^{2}}}{k}\left[ \text{dn}^{-1}\left(
\frac{L}{k}+\alpha ,k\right) -\text{dn}^{-1}\left(
\frac{L}{k}-\alpha ,k\right) \right] , \label{2/35}
\\
\alpha \in \left[ 0,\alpha _{c}\right] , \nonumber
\end{eqnarray}%
for the case (\ref{2/30}) ($p=2m$, $m=0,1,\ldots $), and%
\begin{eqnarray}
H=\frac{1}{2k}\left[ \text{dn}\left( \frac{L}{k}+\alpha ,k\right) +\text{dn}%
\left( \frac{L}{k}-\alpha ,k\right) \right] ,  \label{2/36}
\\
J=\frac{1}{k}\left[ \text{dn}\left( \frac{L}{k}+\alpha ,k\right) -\text{dn}%
\left( \frac{L}{k}-\alpha ,k\right) \right] ,  \label{2/37}
\\
\alpha \in \left[ 0,\alpha _{c}\right] , \nonumber
\end{eqnarray}%
for the case (\ref{2/31}) ($p=2m+1$, $m=0,1,\ldots $), with the
domains of the parameter $k$ being determined by (\ref{2/28}),
(\ref{2/29}). In the
limit $k=1$, equations (\ref{2/34}), (\ref{2/35}) for $p=0$ take the form (%
\ref{2/24}), (\ref{2/25}), as they should.

By setting $\alpha =\alpha _{c}$ in (\ref{2/34})-(\ref{2/37}), we
can obtain relevant parts of the dependence $J_{c}=J_{c}\left(
H\right) $ for arbitrary $L\in \left( 0,\infty \right) $: this is
the subject of the next section. However, we want to conclude this
section by demonstrating how the above exact analytical results
reproduce the well-known\cite{J65,S72,BP82,L86}
"Fraunhofer pattern" of $J_{c}=J_{c}\left( H\right) $ in the limiting case $%
L\equiv \frac{W}{2}\ll 1$.

In the case $L\ll 1$, the solutions to Eqs. (\ref{2/29}) are%
\begin{equation}
k_{p}\thickapprox \frac{2L}{p\pi }\ll 1,\quad p=1,2,\ldots  \label{2/38}
\end{equation}%
(see Fig.~\ref{fig:a&b(k)} for $L=0.3$). Accordingly, the domains
of the parameter $k$
[relations (\ref{2/28})] become%
\begin{equation}
p=0:k\in \left( \frac{2L}{\pi },1\right) ;
\quad p=1,2,\ldots :k\in \left( \frac{2L}{\left( p+1\right) \pi },\frac{2L}{p\pi }%
\right] . \label{2/39}
\end{equation}%
Therefore, we focus our attention on the case $k\ll 1$. In this limit, for $%
k\neq k_{p}$ ($p=1,2,\ldots $), the solution to both Eq. (\ref{2/32}) and \
Eq. (\ref{2/33}) is%
\begin{equation}
\alpha _{c}\thickapprox \frac{\pi }{4}.  \label{2/40}
\end{equation}%
For $k\ll 1$, equations (\ref{2/34}) and (\ref{2/36}) yield%
\begin{equation}
k\thickapprox H^{-1},  \label{2/41}
\end{equation}%
whereas Eqs. (\ref{2/35}), (\ref{2/37}) become%
\begin{equation}
J\thickapprox \left( -1\right) ^{p}\frac{k}{2}\sin \left( \frac{2L}{k}%
\right) \sin \left( 2\alpha \right) ,\quad \alpha \in \left[ 0,\frac{\pi }{4}%
\right] ;  \quad p=0,1,2,\ldots . \label{2/42}
\end{equation}%
Combining relations (\ref{2/39})-(\ref{2/42}), we arrive at an
approximate
dependence $J\thickapprox J\left( H,\alpha \right) $ for $L\equiv \frac{W}{2}%
\ll 1$:%
\begin{equation}
J\thickapprox \frac{1}{2H}\left\vert \sin \left( HW\right) \right\vert \sin
\left( 2\alpha \right) ,\quad \alpha \in \left[ 0,\frac{\pi }{4}\right] .
\label{2/44}
\end{equation}%
As a result,%
\begin{equation}
J_{c}\left( H\right) \thickapprox \frac{1}{2H}\left\vert \sin \left(
HW\right) \right\vert .  \label{2/45}
\end{equation}

Our derivation clearly reveals limitations of the approximate relation (\ref%
{2/45}) (the "Fraunhofer pattern"): strictly speaking, in the field range $%
0<H\lesssim 1$ (i.e., when $k\rightarrow 1$), it can be regarded, at most,
as a reasonable interpolation. Moreover, the approximation (\ref{2/45})
breaks down near the boundaries of the stability regions $H\thickapprox
\frac{p\pi }{W}$, $p=1,2,\ldots $, (i.e., when $k\thickapprox k_{p}$).
Unfortunately, these limitations are not accounted for in elementary
derivations\cite{J65,S72,BP82,L86} of (\ref{2/45}).

Finally, we note that, as is clear from (\ref{2/41}), the actual expansion
parameter in relations (\ref{2/42})-(\ref{2/45}) is $H^{-1}\ll 1$ rather
than $L\ll 1$. Therefore, for $H\gg 1$, the approximation (\ref{2/45}) is
valid for arbitrary $W\in \left( 0,\infty \right) $. (This fact was first
pointed out in Ref. \cite{Zh78}.) For reference purposes, we present the
corresponding ($H\gg 1$) asymptotics of the current-carrying solutions $\phi
_{p}$:%
\begin{eqnarray}
\phi _{p}\left( y\right) \thickapprox p\pi +2Hy+2\alpha -\frac{\alpha }{%
2H^{2}} -\frac{\left( -1\right) ^{p}}{4H^{2}}\left[ \sin \left(
2Hy+2\alpha \right) -2Hy\cos \left( HW\right) \cos \left( 2\alpha
\right) \right] ,  \label{2/46}
\\
\alpha \in \left[ 0,\frac{\pi }{4}\right] ;\quad p=0:H\in \left( 1,\frac{\pi
}{W}\right) ;
\quad p=1,2,\ldots :H\in \left( \frac{p\pi }{W},\frac{\left( p+1\right) \pi }{W}%
\right) . \nonumber
\end{eqnarray}%

\section{Major physical results}

\subsection{Stability regions in the plane $\left( H,J\right) $ and the
dependence $J_{c}=J_{c}\left( H\right) $}

\begin{figure}
\includegraphics{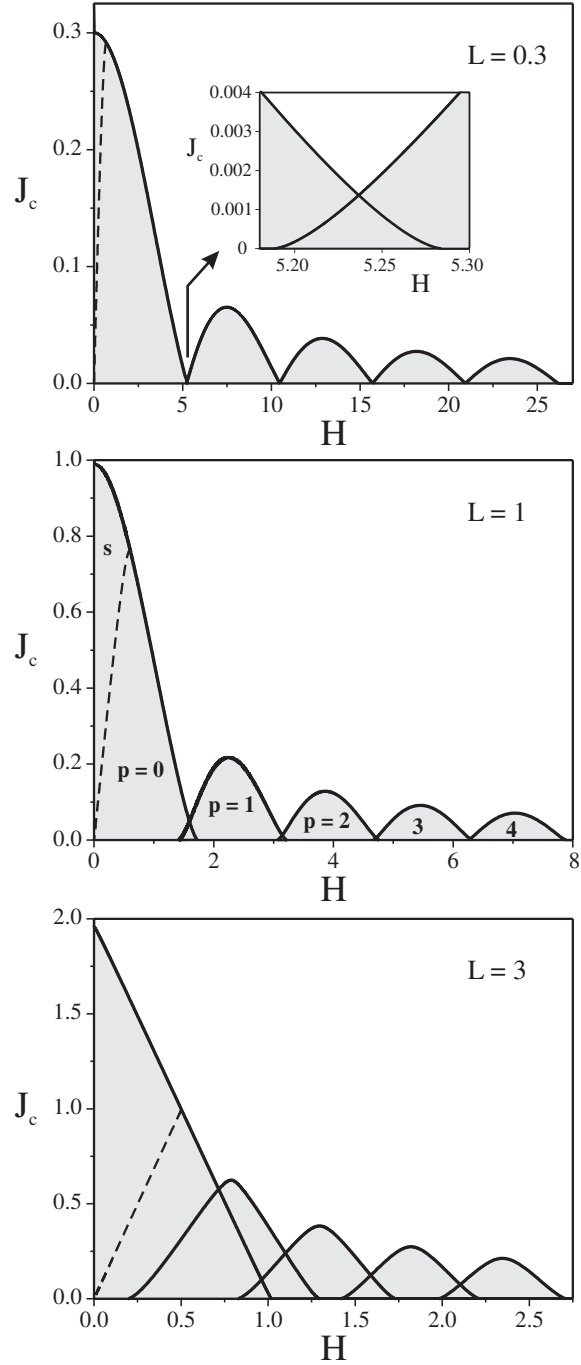}
\caption{\label{fig:Jc(H)all}The stability regions of $\phi _{s}$ and $\phi _{p}$ ($p=0,1,2\ldots $%
) in the physical plane $\left( H,J\right) $ (shaded) for
$L=0.3,1,3$. The dependence $J_{c}=J_{c}\left( H\right) $ is given
by the solid lines. The dashed line represents the internal
boundary where $\phi _{s}=\phi _{0}=\phi _{l\text{.}}$.}
\end{figure}

Relations (\ref{2/22}), (\ref{2/23}) and (\ref{2/34})-(\ref{2/37})
map the
stability regions of the current-carrying solutions (\ref{2/20}) and (\ref%
{2/30}), (\ref{2/31}), respectively, from the parametric planes
$\left( \beta ,k\right) $, $\left( \alpha ,k\right) $ to the
physical plane $\left( H,J\right) $. In Fig.~\ref{fig:Jc(H)all},
we present the results of this mapping for the data of
Fig.~\ref{fig:a&b(k)}. As already noted (Sec. III), the boundaries
of the stability regions represent the dependence
$J_{c}=J_{c}\left( H\right) $ that consists of an infinite number
of separate branches.

As can be easily seen, the structure of the stability regions
[including the boundaries $J_{c}=J_{c}\left( H\right) $] is
qualitatively the same for all the considered cases: $L=0.3$ (a
"small" junction), $L=1$ (a "medium" junction), and $L=3$ (a
"large" junctions). Thus, as the solutions $\phi _{s} $ [Eq.
(\ref{2/20})] and $\phi _{0}$ [Eq. (\ref{2/30})] constitute two
different branches of the same current-carrying solution, their
stability regions (labeled by the indices $s$ and $p=0$,
respectively) merge to form a unified stability domain. The
transformation $\phi _{s}\longleftrightarrow \phi _{0}$ occurs on
the internal boundary (\ref{2/24}), (\ref{2/25}) (represented by
the dashed line in Fig.~\ref{fig:Jc(H)all}), where these two
solutions coincide with the elementary solution $\phi _{l}$ [Eq.
(\ref{2/21.4})].

The stability regions corresponding to $\phi _{p=2m}$ and $\phi _{p=2m+1}$
interchange and form an infinite set. Significantly, for \textit{arbitrary} $%
L\in \left( 0,\infty \right) $, each two consecutive stability regions,
labeled by $p$ and $p+1$, overlap in the field range%
\begin{equation}
\sqrt{H_{p}^{2}-1}\leq H<H_{p},  \label{3/1}
\end{equation}%
where $H_{p}$ are the roots of%
\begin{equation}
\left( p+1\right) K\left( \frac{1}{H_{p}}\right) =H_{p}L,\quad
p=0,1,2,\ldots .  \label{3/2}
\end{equation}%
Indeed, for $J=0$, the stability regions of the solutions $\phi
_{p}$ [Eqs. (\ref{2/26}), (\ref{2/27})] are given by the field
intervals\cite{KG06}
\begin{eqnarray}
p=0:0\leq H<H_{0};  \label{3/3}
\\
p=1,2,\ldots :\sqrt{H_{p-1}^{2}-1}\leq H<H_{p};  \label{3/4}
\end{eqnarray}%
hence relation (\ref{3/1}). Moreover, for sufficiently large $L$,
the
overlap may involve several consecutive stability regions: see Fig.~\ref{fig:Jc(H)all} for $%
L=3$. In contrast, the overlap decreases with an increase in $p$
and a
decrease in $L$: see the insert in Fig.~\ref{fig:Jc(H)all} for $%
L=0.3$. The overlap of the stability regions results in
multivaluedness of the dependence $J_{c}=J_{c}\left( H\right) $.

For $L\gtrsim 1$, the multivaluedness of $J_{c}=J_{c}\left( H\right) $ was
found by numerical evaluation.\cite{OS67} However, the fact that this
multivaluedness is an intrinsic feature of any Josephson junction (even with
$L\ll 1$) was not noticed because of the absence of exact analytical
solutions.

\subsection{Unquantized Josephson vortices}

In contrast to the case $J=0$,\cite{r3} the discrete parameter $p$ of the
exact solutions (\ref{2/30}) and (\ref{2/31}) for $J>0$ cannot be identified
with the number of Josephson vortices, although relation (\ref{2/29.1})
still holds. The reason is the occurrence (for certain values of $H$ and $J$%
) of \textit{unquantized} vortices carrying fractional flux $\Phi
\in \left( \frac{1}{2}\Phi _{0},\Phi _{0}\right) $, where $\Phi
_{0}$ is the flux quantum. (In our dimensionless units, $\Phi
_{0}=\pi $.) To clarify the situation, we should consider spatial
distribution of the local magnetic field $h$ and of the Josephson
current density $j$ for $J=J_{c}$.

As follows from (\ref{0/1}) and (\ref{1/3}), the local magnetic field $h$
obeys the linear homogeneous second-order differential equation%
\begin{equation}
\frac{d^{2}h}{dy^{2}}=\cos \phi \left( y\right) h.  \label{3/5}
\end{equation}%
Combining Eq. (\ref{3/5}) and Eqs. (\ref{1/11}), (\ref{1/12}) for the
boundary of the stability region, we obtain%
\begin{equation}
\bar{\psi}_{0}\left( L\right) \frac{dh}{dy}\left( L\right) =\bar{\psi}%
_{0}\left( -L\right) \frac{dh}{dy}\left( -L\right) .  \label{3/6}
\end{equation}%
Taking into account (\ref{1/13}), we find that either%
\begin{equation}
\frac{dh}{dy}\left( L\right) =\frac{dh}{dy}\left( -L\right) =0,  \label{3/7}
\end{equation}%
or%
\begin{equation}
\frac{dh}{dy}\left( L\right) \frac{dh}{dy}\left( -L\right) >0.  \label{3/8}
\end{equation}

The solution $\phi _{s}$ [Eq. (\ref{2/20})] satisfies relation (\ref{3/8})
everywhere on the critical curve $J_{c}=J_{c}\left( H\right) $. Moreover,
for this solution, $\frac{dh}{dy}\left( y\right) >0$ for any $y\in \left[
-L,L\right] $. [Accordingly, $j\left( y\right) >0$ for any $y\in \left[ -L,L%
\right] $: see (\ref{0/2}) and (\ref{1/3}).]

The behavior of $\phi _{p}$ [Eqs. (\ref{2/30}) and (\ref{2/31})] is more
complicated. First, we note that $\phi _{p}$ satisfy (\ref{3/7}) at those
values of $H$ for which $J_{c}=0$. This occurs at $H=H_{0}$ (for $\phi _{0}$%
) and at $H=\sqrt{H_{p-1}^{2}-1},H_{p}$ (for $\phi _{p}$with $p=1,2,\ldots $%
), where $H_{p}$ are determined by (\ref{3/2}): as a matter of fact, this
case has been considered in detail in Ref. \cite{KG06}.

In addition, relation (\ref{3/7}) is satisfied by $\phi _{p}$ at such fields
$H=H_{p}^{\ast }$ ($p=0,1,2,\ldots $) that $J_{c}=\left( 2H_{p}^{\ast
}\right) ^{-1}$. These fields are given by%
\[
H_{p}^{\ast }=\frac{1}{2k_{p}^{\ast }}\left[ 1+\sqrt{1-\left( k_{p}^{\ast
}\right) ^{2}}\right] ,\quad p=0,1,2,\ldots ,
\]%
where $k_{p}^{\ast }$ are determined by (\ref{2/33.3}). At $H=H_{p}^{\ast }$%
, we have: $\phi _{p}\left( -L\right) =0$, $\phi _{p}\left( L\right) =\pi
\left( 2p+1\right) $.

At the rest of the points on the critical curves
$J_{c}=J_{c}\left( H\right) $, the solutions $\phi _{p}$
($p=0,1,2,\ldots $) satisfy (\ref{3/8}).
In particular, we have: (a) $\frac{dh}{dy}\left( \pm L\right) >0$ for $%
H<H_{p}^{\ast }$ ($p=0,1,2,\ldots $), because $\phi _{p}\left(
-L\right) \in \left( 0,\frac{\pi }{2}\right) $ and $\phi
_{p}\left( L\right) \in \left( 2\pi p,2\pi \left(
p+\frac{1}{2}\right) \right) $; (b) $\frac{dh}{dy}\left( \pm
L\right) <0$ for $H>H_{p}^{\ast }$ ($p=0,1,2,\ldots $), because
$\phi _{p}\left( -L\right) \in \left( -\pi ,0\right) $ and $\phi
_{p}\left( L\right) \in \left( 2\pi \left( p+\frac{1}{2}\right)
,2\pi \left( p+1\right) \right) $.

To establish the types of Josephson-vortex structures that are represented
by the solutions $\phi _{p}$ ($p=0,1,2,\ldots $) on the critical curve $%
J_{c}=J_{c}\left( H\right) >0$, we have to classify the points of local
minima of $h$. Thus, for $H<H_{p}^{\ast }$ ($p=0,1,2,\ldots $), the first
minimum is positioned at $y=-L$, where $\frac{dh}{dy}\left( -L\right) >0$.
The rest of the minima (for $p>0$) are positioned at $y=y_{n}$ ($n=1,\ldots
,p$), where $\phi _{p}\left( y_{n}\right) =2\pi n$ and $\frac{dh}{dy}\left(
y_{n}\right) =0$.

For $H=H_{p}^{\ast }$ ($p=0,1,2,\ldots $), the first minimum is positioned
at $y=-L$, where $\phi _{p}\left( -L\right) =0$ and $\frac{dh}{dy}\left(
-L\right) =0$. The rest of the minima (For $p>0$) are positioned at $y=y_{n}$
($n=1,\ldots ,p$), where $\phi _{p}\left( y_{n}\right) =2\pi n$ and $\frac{dh%
}{dy}\left( y_{n}\right) =0$.

For $H>H_{p}^{\ast }$ ($p=0,1,2,\ldots $), we have a minimum at $y=L$, where
$\frac{dh}{dy}\left( L\right) <0$. The rest of the minima are positioned at $%
y=y_{n}$ ($n=0,\ldots ,p$), where $\phi _{p}\left( y_{n}\right) =2\pi n$ and
$\frac{dh}{dy}\left( y_{n}\right) =0$.

Bearing in mind that a Josephson vortex is located between two consecutive
local minima of $h$,\cite{KG06} we arrive at the following physical
interpretation of $\phi _{p}$:

(i) $H<H_{p}^{\ast }$ ($p=0,1,2,\ldots $). The solution $\phi _{0}$
represents a vortex-free configuration. The solutions labeled by $%
p=1,2,\ldots $ represent configurations with $p-1$ quantized Josephson
vortices located between the points $y_{n}$, $y_{n+1}$ ($n=1,\ldots ,p-1$)
and carrying flux $\Phi =\Phi _{0}$. In addition, these configurations
contain a single \textit{unquantized }vortex carrying flux $\Phi \in \left(
\frac{1}{2}\Phi _{0},\Phi _{0}\right) $ and located between $y=-L$ and $%
y=y_{1}$;

(ii) $H=H_{p}^{\ast }$ ($p=0,1,2,\ldots $). The solution $\phi _{0}$
represents a vortex-free configuration. The solutions labeled by $%
p=1,2,\ldots $ represent configurations with $p$ quantized Josephson
vortices located between the points $y=-L$, $y_{1}$, and $y_{n}$, $y_{n+1}$ (%
$n=1,\ldots ,p-1$; $p>1$);

(iii) $H>H_{p}^{\ast }$ ($p=0,1,2,\ldots $). The solutions labeled by $%
p=1,2,\ldots $ represent configurations with $p$ quantized Josephson
vortices located between the points $y_{n}$, $y_{n+1}$ ($n=0,\ldots ,p-1$).
\ In addition, all these configurations ($p=0,1,2,\ldots $) contain a single
\textit{unquantized }vortex located between $y_{p}$ and $y=L$.

\begin{figure}
\includegraphics{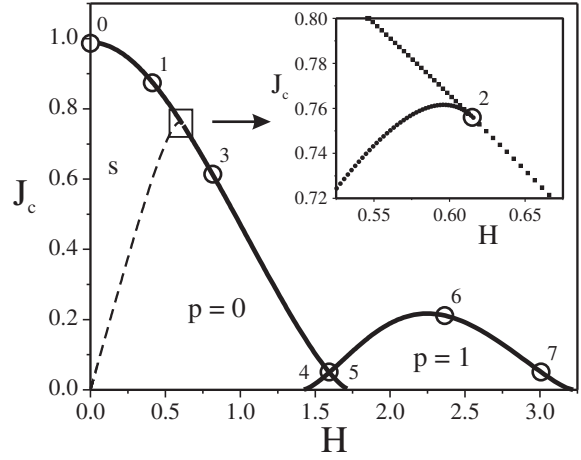}
\caption{\label{fig:Jc(H)}The first two critical curves
$J_{c}=J_{c}\left( H\right) $ for $L=1$. Spatial distribution of
$h$ and $j$ (Fig.~\ref{fig:h&j(y)}) is evaluated at typical points
0-7: (0) $H=0.00$, $J_{c}=0.99$; (1) $H=0.43$, $J_{c}=0.87$; (2) $H=0.61$, $%
J_{c}=0.76$ ($\phi _{s}=\phi _{0}=\phi _{l\text{.}}$); (3)
$H=H_{0}^{\ast
}=0.81$, $J_{c}=\left( 2H_{0}^{\ast }\right) ^{-1}=0.61$; (4) $H=1.60$, $%
J_{c}=0.05$ (the first curve); (5) $H=1.60$, $J_{c}=0.05$ (the
second curve); (6) $H=H_{1}^{\ast }=2.36$, $J_{c}=\left(
2H_{1}^{\ast }\right) ^{-1}=0.21$; (7) $H=3.01$, $J_{c}=0.05$.}
\end{figure}

\begin{figure*}
\includegraphics{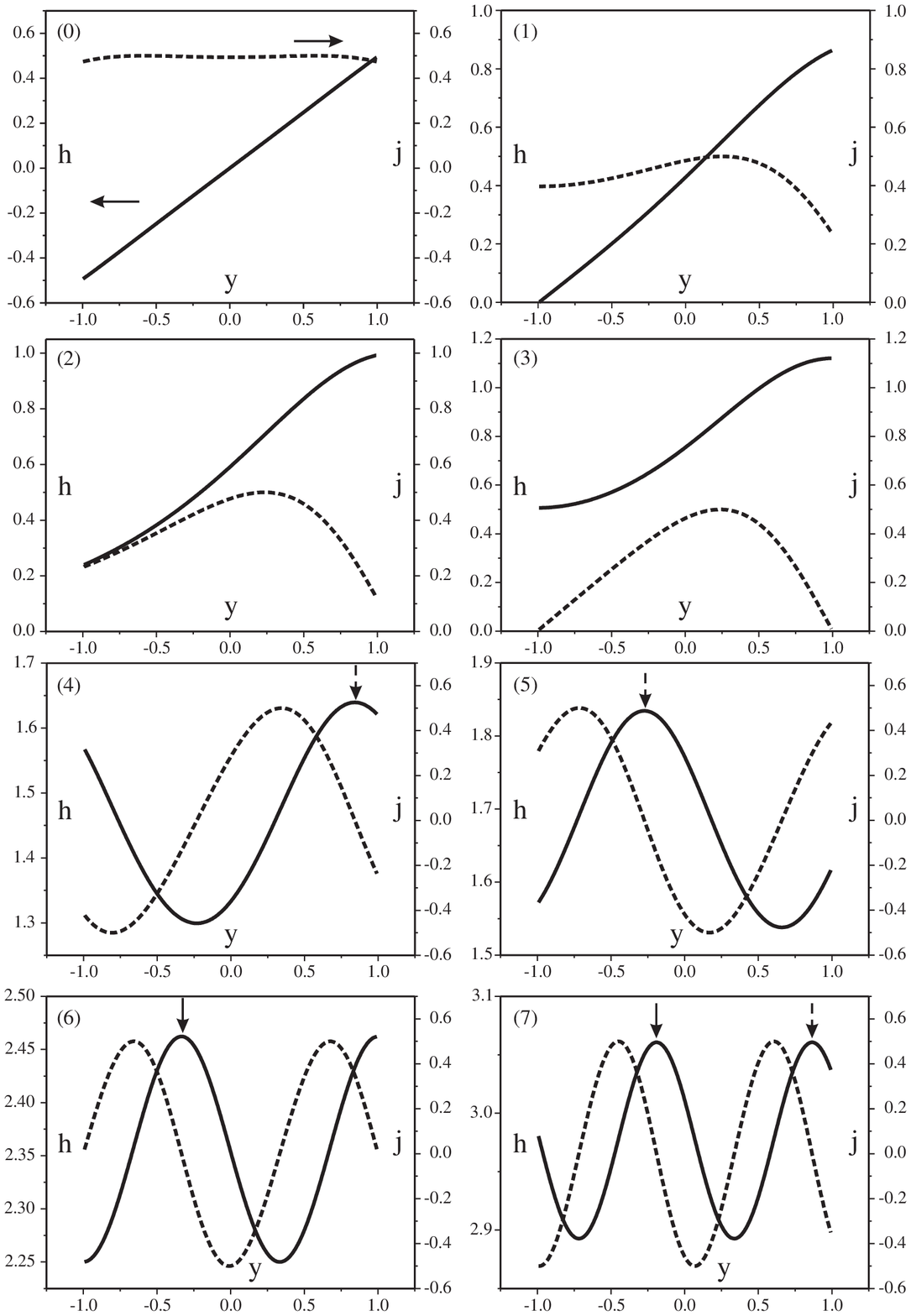}
\caption{\label{fig:h&j(y)}Spatial distribution of $h$ (solid
line) and $j$ (dashed line) for points 0-7 in
Fig.~\ref{fig:Jc(H)}. The location of Josephson vortices is marked
by vertical arrows: the dashed arrows correspond to unquantized
vortices [figures (4), (5) and (7)]; the solid arrows correspond
to quantized vortices [figures (6) and (7)].}
\end{figure*}

The above general analytical conclusions are illustrated in
Figs.~\ref{fig:Jc(H)} and \ref{fig:h&j(y)}. For simplicity, in
Fig.~\ref{fig:Jc(H)}, we restrict ourselves to the first two
critical curves $J_{c}=J_{c}\left( H\right) $ of the junction with
$L=1$. Spatial distribution of $h$ and $j$ at typical points 0-7
on these curves is presented in Fig.~\ref{fig:h&j(y)}, were we
also mark the locations of both quantized and unquantized
Josephson vortices.

In conclusion, we want to emphasize that, as follows from continuity
arguments, unquantized Josephson vortices persist in certain two-dimensional
domains on the plane $\left( H,J\right) $, where $J<J_{c}$. Therefore, the
existence of such vortices is a typical feature of any Josephson junction in
the presence of externally applied magnetic fields and transport currents.

\subsection{Generalizations}

The restriction $H\geq 0$, $J\geq 0$ imposed at the beginning of Sec. III
can be easily removed. Physical solutions that do not obey these restriction
are expressed via the solutions $\phi _{s}$, $\phi _{p}$ and $\phi _{l}$
[Eqs. (\ref{2/20}), (\ref{2/30}), (\ref{2/31}), and (\ref{2/21.4}),
respectively] by means of elementary symmetry relations.

\textit{1. The case }$H\leq 0$\textit{, }$J\geq 0$\textit{:}%
\begin{equation}
\phi _{s},\beta \rightarrow \phi _{s},-\beta ;\qquad
\phi_{p},\alpha \rightarrow -\phi _{p},-\alpha ;\qquad
\phi_{l},\gamma \rightarrow 2\pi -\phi _{l},-\gamma .
\label{4/1}%
\end{equation}

\textit{2. The case }$H\geq 0$\textit{, }$J\leq 0$\textit{:}%
\begin{equation}
\phi _{s},\beta \rightarrow -\phi _{s},-\beta ;\qquad
\phi_{p},\alpha\rightarrow \phi _{p},-\alpha ;\qquad
\phi_{l},\gamma\rightarrow \phi _{l}-2\pi ,-\gamma .
\label{4/2}%
\end{equation}

\textit{3. The case }$H\leq 0$\textit{, }$J\leq 0$\textit{:}%
\begin{equation}
\phi _{s},\beta \rightarrow -\phi _{s},\beta ;\qquad
\phi_{p},\alpha \rightarrow -\phi _{p},\alpha ;\qquad
\phi_{l},\gamma \rightarrow -\phi _{l},\gamma .
\label{4/3}%
\end{equation}

\begin{figure}
\includegraphics{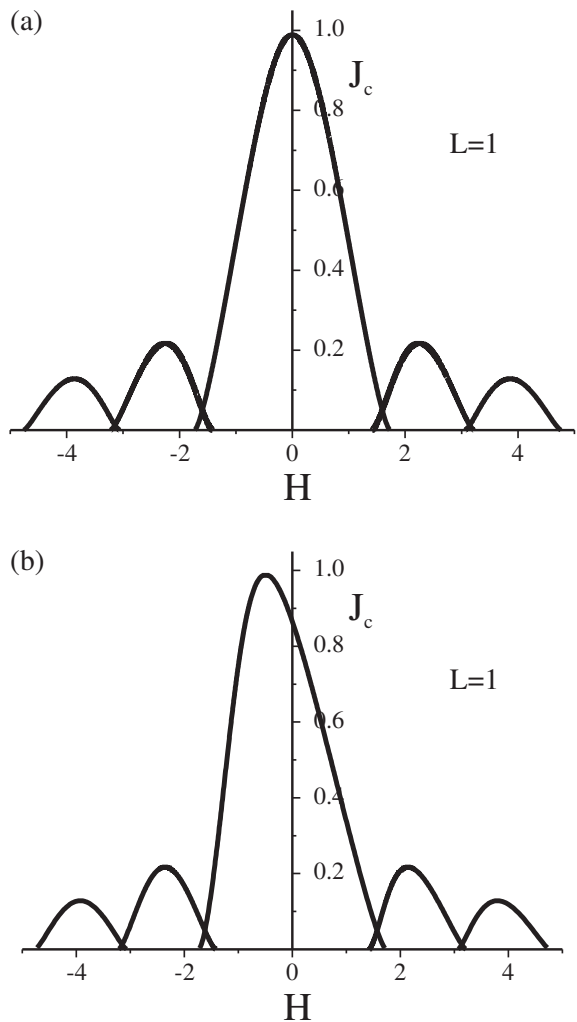}
\caption{\label{fig:Jc(H)asym}The effect of asymmetric injection
of the transport current on the dependence $J_{c}=J_{c}\left(
H\right) $ for $L=1$: (a) $a_{\pm }=\frac{1}{2} $; (b) $a_{+}=0$,
$a_{-}=1$.}
\end{figure}

Finally, we note that the consideration of this paper equally applies to a
generalized form of the boundary conditions that takes into account possible
asymmetry in the injection of the transport current, namely,\cite{BP82,L86}%
\begin{equation}
\frac{d\phi }{dy}\left( \pm L\right) =2\left( H\pm a_{\mp
}J\right) , \label{4/4}
\end{equation}%
or, equivalently,%
\begin{eqnarray}
H=\frac{1}{2}\left[ a_{+}\frac{d\phi }{dy}\left( +L\right)
+a_{-}\frac{d\phi }{dy}\left( -L\right) \right] , \nonumber
\\
J=\frac{1}{2}\left[ \frac{d\phi }{dy}\left( +L\right) -\frac{d\phi }{dy}%
\left( -L\right) \right] . \nonumber
\end{eqnarray}%
where%
\[
a_{\pm }\geq 0,\quad a_{-}+a_{+}=1.
\]%
The effect of the generalized boundary conditions (\ref{4/4}) is
illustrated in Fig.~\ref{fig:Jc(H)asym}.

\section{Summary and conclusions}

Summarizing, we have derived the complete set of exact physical solutions to
the general boundary-value problem (\ref{1/3}), (\ref{4/4}): $\phi _{s}$
[Eq. (\ref{2/20})] and $\phi _{p}$ ($p=0,1,2,\ldots $) [Eqs. (\ref{2/30}), (%
\ref{2/31})]\ complemented by the symmetry relations (\ref{4/1})-(\ref{4/3}%
). The obtained solutions describe the current-carrying states of the
Josephson junction of arbitrary length $W\equiv 2L\in \left( 0,\infty
\right) $ in the presence of an externally applied magnetic field $H\in
\left( -H_{c},H_{c}\right) $, where $H_{c}$ is the thermodynamic critical
field of the superconducting electrodes. The most direct application of
these solutions is straightforward evaluation of the dependence $%
J_{c}=J_{c}\left( H\right) $ (for arbitrary $W$ and an arbitrary mode of the
injection of $J$) by means of the algorithm of Secs. III and IV.

Mathematically, $\phi _{s}$ and $\phi _{p}$ ($p=0,1,2,\ldots $) represent
the complete set of particular solutions to (\ref{1/3}) that are stable
under the condition that $\frac{d\phi }{dy}$ is fixed at the boundaries $%
y=\pm L$, and they possess a number of interesting properties. For example,
the solutions $\phi _{s}$ and $\phi _{0}$ constitute two different branches
of the same stable solution: for $k\rightarrow 1$, both of them turn into
the elementary solution $\phi _{l}$ [Eq. (\ref{2/21.4})]. In physical
literature,\cite{S72,BP82,L86} the elementary solution $\phi _{l}$ is
usually identified with "the Josephson vortex". Indeed, if Eq. (\ref{1/3})
were considered on the \textit{infinite} interval $\left( -\infty ,\infty
\right) $, this solution would be nothing but the well-known\cite{DEGM82}
static soliton of the sine-Gordon equation, positioned at $y=\gamma $ and
stable for arbitrary $\left\vert \gamma \right\vert <\infty $. However, on
the physically realistic \textit{finite} interval $\left[ -L,L\right] $, the
solution $\phi _{l}$ proves to be stable only for $\left\vert \gamma
\right\vert \geq \gamma _{c}$, where $\gamma _{c}$ is determined by Eq. (\ref%
{2/21.2}), and, as shown in Sec. IV, it has nothing to do with any vortex
(or soliton) configurations.

As could be anticipated, the exact analytical solution of the problem that
remained unresolved for over four decades has revealed some unexpected
physical features. For example, contrary to a wide-spread belief,\cite%
{S72,BP82,L86} it clearly shows that there is no \textit{qualitative}
difference between Josephson junctions with $W\gg 1$ and those with $W\ll 1$%
. Thus, the \textit{exact} analytical dependence $J_{c}=J_{c}\left( H\right)
$ proves to be multivalued even for arbitrarily small $W$. Therefore,
hysteresis is an intrinsic feature of any Josephson junction with $W\in
\left( 0,\infty \right) $.

However, we think that the most important physical conclusion that can be
drawn from the exact analytical solution is the existence of \textit{%
unquantized} Josephson vortices. Indeed, recently, the possibility of
finding unquantized vortices in different types of superconducting systems
(including Josephson ones) has attracted considerable attention: see, e.g.,
Refs. \cite{B02,Mi02} \ and references therein. In most theoretical models,
unquantized vortices appear as a result of unconventional properties of the
superconductors themselves, such as, e.g., the existence of two
superconducting order parameters,\cite{B02} $d$-wave pairing combined with
the inhomogeneity of grain boundaries,\cite{Mi02} etc. By contrast, we have
shown that the presence of unquantized Josephson vortices near the external
boundaries is a typical feature of any classical Josephson junction,
provided the transport current $J$ is sufficiently close to $J_{c}$ for
certain finite values of $H$.

From a mathematical point of view, it would be desirable to know whether the
quantity $J_{\max }$ discussed in the Introduction indeed coincides with $%
J_{c}$ evaluated in this paper. Although we have been unable to find a
general analytical proof, our detailed comparisons with the numerical
results of Refs. \cite{OS67,BB75} (not presented here for brevity reasons)
imply that the identity $J_{\max }\equiv J_{c}$ can be accepted without
reservation.

Finally, we want to remind once again that Eq. (\ref{1/3}) is just
the
static version of the well-known sine-Gordon equation [Eq. (\ref{a1}) with $%
\kappa =0$]. Given that the sine-Gordon equation finds a lot of applications
in condensed-matter and elementary-particle physics,\cite{DEGM82} we hope
that our exact analytical solution may find applications outside the field
of superconductivity as well.

\begin{center}
\bigskip\ \textbf{Acknowledgements}
\end{center}

We thank A. N. Omelyanchouk, A. S. Kovalev, and M. M. Bogdan for stimulating
discussions of the main physical and mathematical results of the paper.

\appendix

\section{Alternative formulation of the stability problem}

The stability of a given solution $\phi =\phi \left( y\right) $ to (\ref{1/3}%
), (\ref{1/4}) can also be analyzed using the general time-dependent equation%
\cite{BP82}%
\begin{equation}
\frac{\partial ^{2}\phi }{\partial t^{2}}+2\kappa \frac{\partial \phi }{%
\partial t}-\frac{\partial ^{2}\phi }{\partial y^{2}}+\sin \phi =0,
\label{a1}
\end{equation}%
where $\kappa >0$, $t\geq 0$ and $y\in \left( -L,L\right) $, under the
boundary conditions%
\begin{equation}
\frac{d\phi }{dt}\left( t,\pm L\right) =2H\pm J.  \label{a2}
\end{equation}

According to linear stability theory,\cite{JJ80} we should seek solutions to
(\ref{a1}), (\ref{a2}) in the form%
\begin{equation}
\tilde{\phi}\left( t,y\right) =\phi \left( y\right) +e^{-\sigma t}\zeta
\left( y\right) ,  \label{a3}
\end{equation}%
where%
\[
\max_{y}\left\vert \zeta \left( y\right) \right\vert \ll 1,
\]%
and
\begin{equation}
\frac{d\zeta }{dy}\left( \pm L\right) =0.  \label{a4}
\end{equation}
Substituting (\ref{a3}) into (\ref{a1}) and dropping nonlinear terms, we
obtain:%
\begin{equation}
-\frac{d^{2}\zeta }{dy^{2}}+\cos \phi \left( y\right) \zeta =\sigma \left(
2\kappa -\sigma \right) \zeta .  \label{a5}
\end{equation}

Equation (\ref{a5}) under boundary conditions (\ref{a4}) immediately yields%
\begin{equation}
\sigma _{n\pm }=\kappa \pm \sqrt{\kappa ^{2}-\mu _{n}},\quad n=0,1,2,\ldots ,
\label{a6}
\end{equation}%
where $\mu _{0}<\mu _{1}<\mu _{2}<\ldots $ are the eigenvalues of the
Sturm-Liouville problem (\ref{1/7}), (\ref{1/8}). Thus, we arrive at the
following classification of stability properties of the solution $\phi =\phi
\left( y\right) $:

i) $\mu _{0}>0$, $\text{Re }\sigma _{n\pm }>0$ ($n=0,1,2,\ldots
$): the solution is \textit{exponentially} stable;

ii) $\mu _{0}<0$, $\sigma _{0-}=\kappa -\sqrt{\kappa ^{2}+\left\vert \mu
_{0}\right\vert }<0$: the solution is unstable;

iii) $\mu _{0}=0$, $\sigma _{0-}=0$: the solution is at the boundary of the
stability region.

\section{Solution of the linear boundary-value problem for $\protect\mu %
_{0}=0$, $\protect\psi _{0}=\bar{\protect\psi}_{0}$}

To solve the linear boundary-value problem (\ref{1/11})-(\ref{1/13}), we
should first find the general solution to (\ref{1/11}). It can be written in
the form%
\begin{equation}
\bar{\psi}_{0}\left( y\right) =C_{1}\chi _{1}\left( y\right) +C_{2}\chi
_{2}\left( y\right) ,  \label{b4}
\end{equation}%
where $\chi _{1}$, $\chi _{2}$ are linearly independent solutions to (\ref%
{1/11}), and $C_{1}$, $C_{2}$ are arbitrary constants. As to $\chi _{1}$, we
can choose\cite{KG06}%
\begin{equation}
\chi _{1}=\tilde{C}\frac{d\phi }{dy},  \label{b5}
\end{equation}%
where $\tilde{C}$ is a normalization constant. The linearly
independent
solution $\chi _{2}$ is determined by the well-known relation\cite{S64}%
\begin{equation}
\chi _{2}=\chi _{1}\int \frac{dy}{\chi _{1}^{2}}.  \label{b6}
\end{equation}

In the simplest situations, we have either%
\begin{equation}
\chi _{1}\left( -y\right) =\chi _{1}\left( y\right) ,\quad \chi _{2}\left(
-y\right) =-\chi _{2}\left( y\right) ,  \label{b7}
\end{equation}%
or%
\begin{equation}
\chi _{1}\left( -y\right) =-\chi _{1}\left( y\right) ,\quad \chi _{2}\left(
-y\right) =\chi _{2}\left( y\right) .  \label{b8}
\end{equation}%
In the case (\ref{b7}), which corresponds to $\phi \equiv \phi _{p}$ [Eqs. (%
\ref{2/26}), (\ref{2/27})] with $k=k_{p}$, we have $C_{2}=0$, $\bar{\psi}%
_{0}=\chi _{1}$. Conditions (\ref{1/12}) result in Eqs. (\ref{2/29}).\cite%
{KG06} Condition (\ref{1/13}) is fulfilled automatically.

In the case (\ref{b8}), which corresponds to $\phi \equiv \phi _{s}$ [Eqs. (%
\ref{2/9})] with $k=k_{c}$, we have $C_{1}=0$,%
\begin{equation}
\bar{\psi}_{0}\left( y\right) =\chi _{2}\left( y\right) \equiv
\frac{\text{sn}\left( y,k_{c}\right) }{\text{dn}\left(
y,k_{c}\right)
}\left[ -E\left( y,k_{c}\right) +\left( 1-k_{c}^{2}\right) y\right] -\text{cn%
}\left( y,k_{c}\right) .  \label{b10}
\end{equation}%
The substitution of (\ref{b10}) into (\ref{1/12}) yields Eq.
(\ref{2/11}). Condition (\ref{1/13}) singles out the solution
presented in Fig.~\ref{fig:kc(L)}.

Consider now the general situation, when the functions $\chi _{1}$, $\chi
_{2}$ do not obey either (\ref{b7}) or (\ref{b8}), and, accordingly, $%
C_{1}\neq 0$, $C_{2}\neq 0$. Upon the substitution of (\ref{b4}) into (\ref{1/12}%
), we get a system of algebraic equations for $C_{1}$, $C_{2}$,%
\begin{eqnarray}
C_{1}\frac{d\chi _{1}}{dy}\left( L\right) +C_{2}\frac{d\chi
_{2}}{dy}\left( L\right) =0, \nonumber
\\
C_{1}\frac{d\chi _{1}}{dy}\left( -L\right) +C_{2}\frac{d\chi
_{2}}{dy}\left( -L\right) =0, \nonumber
\end{eqnarray}%
with%
\begin{equation}
\frac{d\chi _{1}}{dy}\left( L\right) \frac{d\chi _{2}}{dy}\left( -L\right) =%
\frac{d\chi _{1}}{dy}\left( -L\right) \frac{d\chi _{2}}{dy}\left( L\right)
\label{b12}
\end{equation}%
being the solvability condition. Equation (\ref{b12}), under condition (\ref%
{1/13}), determines the sought solution.

In particular, in the case of (\ref{2/20}) with $\beta =\beta _{c}$, we have%
\begin{eqnarray}
\chi _{1}\left( y\right) =\frac{\text{sn}\left( y+\beta _{c},k\right) }{%
\text{dn}\left( y+\beta _{c},k\right) },  \label{b14}
\\
\chi _{2}\left( y\right) =\frac{\text{sn}\left( y+\beta _{c},k\right) }{%
\text{dn}\left( y+\beta _{c},k\right) }\left[ -E\left( y+\beta
_{c},k\right) +\left( 1-k^{2}\right) y\right] -\text{cn}\left(
y+\beta _{c},k\right) .  \label{b15}
\end{eqnarray}%
The substitution of (\ref{b14}), (\ref{b15}) into (\ref{b12}) leads to Eq. (%
\ref{2/21}). Condition (\ref{1/13}) leads to the boundary condition $\beta
_{c}\left( k_{c}\right) =0$ for the domain (\ref{2/10}), where $k_{c}$ is
determined by Eq. (\ref{2/11}).

In the case of (\ref{2/30}) with $\alpha =\alpha _{c}$, the functions $\chi
_{1}$, $\chi _{2}$ are given by%
\begin{eqnarray}
\chi _{1}\left( y\right) =\text{dn}^{-1}\left( \frac{y}{k}+\alpha
_{c},k\right) ,  \label{b18}
\\
\chi _{2}\left( y\right) =\text{dn}^{-1}\left( \frac{y}{k}+\alpha
_{c},k\right) E\left( \frac{y}{k}+\alpha _{c},k\right) .
\label{b19}
\end{eqnarray}%
Substituting (\ref{b18}) and (\ref{b19}) into (\ref{b12}), we get Eq. (\ref%
{2/32}). Analogously, for (\ref{2/31}) with $\alpha =\alpha _{c}$,
\begin{eqnarray}
\chi _{1}\left( y\right) =\text{dn}\left( \frac{y}{k}+\alpha
_{c},k\right) , \label{b22}
\\
\chi _{2}\left( y\right) =\frac{\text{dn}\left( \frac{y}{k}+\alpha
_{c},k\right) }{1-k^{2}}
\left[ E\left( \frac{y}{k}+\alpha _{c},k\right) -\frac{k^{2}%
\text{sn}\left( \frac{y}{k}+\alpha _{c},k\right) \text{cn}\left( \frac{y}{k}%
+\alpha _{c},k\right) }{\text{dn}\left( \frac{y}{k}+\alpha _{c},k\right) }%
\right] ,  \label{b23}
\end{eqnarray}%
with Eq. (\ref{2/33}) being the result of substitution into
(\ref{b12}). Condition (\ref{1/13}) leads to the boundary
conditions $\alpha _{c}\left(
k_{p}\right) =0$ ($p=1,2,\ldots $) for the domains (\ref{2/28}), where $%
k_{p} $ are determined by Eqs. (\ref{2/29}).

Finally, in the case of the elementary solution (\ref{2/21.4}) with $\gamma
=\gamma _{c}$,%
\begin{eqnarray}
\chi _{1}\left( y\right) =\cosh ^{-1}\left( y-\gamma _{c}\right) ,
\label{b24}
\\
\chi _{2}\left( y\right) =\frac{1}{2\cosh \left( y-\gamma _{c}\right) }\left[
\frac{\sinh \left( y-\gamma _{c}\right) }{2}+y\right] .  \label{b25}
\end{eqnarray}%
Substituting (\ref{b24}), (\ref{b25}) into (\ref{b12}), we arrive at Eq. (%
\ref{2/21.1}) that has been obtained in the main text as the limiting form
of Eqs. (\ref{2/21}) and (\ref{2/32}) (with $p=0)$ for $k\rightarrow 1$.

\section{Explicit evaluation of $\protect\mu =\protect\mu _{0}$ for $H\gg 1$}

For the lowest eigenvalue $\mu =\mu _{0}$, the Sturm-Liouville problem (\ref%
{1/7}), (\ref{1/8}) becomes%
\begin{eqnarray}
-\frac{d^{2}\psi _{0}}{dy^{2}}+\cos \phi \left( y\right) \psi
_{0}=\mu _{0}\psi _{0},\quad y\in \left( -L,L\right) ,  \label{c1}
\\
\frac{d\psi _{0}}{dy}\left( -L\right) =\frac{d\psi _{0}}{dy}\left( L\right)
=0,  \label{c2}
\\
\psi _{0}\left( y\right) \neq 0,\quad y\in \left[ -L,L\right] .  \label{c3}
\end{eqnarray}%
In the general case, the solution to (\ref{c1})-(\ref{c3}) can be
obtained
using the fact that Eq. (\ref{c1}) is reducible to Lam\'{e}'s equation.\cite%
{WW27} However, for $H\gg 1$, the eigenvalue $\mu _{0}$ can be explicitly
evaluated by elementary methods.

We will seek the solution to (\ref{c1})-(\ref{c3}) in the form of asymptotic
expansions%
\begin{equation}
\psi _{0}\left( y\right) \thickapprox \sum_{n\geq 0}\psi _{0}^{\left(
n\right) }\left( y\right) ,\quad \mu _{0}\thickapprox \sum_{n\geq 1}\mu
_{0}^{\left( n\right) },  \label{c4}
\end{equation}%
where $\psi _{0}^{\left( n\right) }$ and $\mu _{0}^{\left( n\right) }$ are
of order $H^{-n}$. \ Besides, we will employ the exact integral relation
\begin{equation}
\mu _{0}=\frac{\int_{-L}^{L}dy\psi _{0}\left( y\right) \cos \phi \left(
y\right) }{\int_{-L}^{L}dy\psi _{0}\left( y\right) }  \label{c5}
\end{equation}%
that immediately follows from (\ref{c1})-(\ref{c3}).

Introducing a new variable $u\equiv Hy$, we rewrite (\ref{c1})-(\ref{c3}) as%
\begin{eqnarray}
-\frac{d^{2}\psi _{0}}{du^{2}}+\frac{1}{H^{2}}\left[ \cos \phi
\left( u\right) -\mu _{0}\right] \psi _{0}=0,\quad u\in \left(
-HL,HL\right) , \label{c6}
\\
\frac{d\psi _{0}}{du}\left( -HL\right) =\frac{d\psi _{0}}{du}\left(
HL\right) =0,  \label{c7}
\\
\psi _{0}\left( u\right) \neq 0,\quad u\in \left[ -HL,HL\right]  \label{c8}
\end{eqnarray}%
and note that $\left\vert \mu _{0}\right\vert \leq 1$.\cite{KG06}
Thus, the
problem for $\psi _{0}^{\left( 0\right) }$ has the form%
\begin{eqnarray}
-\frac{d^{2}\psi _{0}^{\left( 0\right) }}{du^{2}}=0,\quad u\in
\left( -HL,HL\right) ,  \label{c9}
\\
\frac{d\psi _{0}^{\left( 0\right) }}{du}\left( -HL\right) =\frac{d\psi
_{0}^{\left( 0\right) }}{du}\left( HL\right) =0,  \label{c10}
\\
\psi _{0}^{\left( 0\right) }\left( u\right) \neq 0,\quad u\in \left[ -HL,HL%
\right] .  \label{c11}
\end{eqnarray}%
The solution to (\ref{c9})-(\ref{c11}) is%
\begin{equation}
\psi _{0}^{\left( 0\right) }\left( y\right) =\text{const}.
\label{c12}
\end{equation}%
Using relation (\ref{c5}), solution (\ref{c12}) and the asymptotic
expansions
for $\phi _{p}$ [relation (\ref{2/46})], we find%
\begin{equation}
\mu _{0}\left( H,\alpha \right) \thickapprox \mu _{0}^{\left( 1\right)
}\left( H,\alpha \right) =\frac{\left( -1\right) ^{p}}{HW}\sin \left(
HW\right) \cos \left( 2\alpha \right) .  \label{c13}
\end{equation}%
As can be easily seen, expression (\ref{c13}) stands in full agreement with
the general results of Sec. III.


\begin{thebibliography}{99}
\bibitem{r2} This follows, e.g., from the fact that the Sturm-Liouville
problem (\ref{1/7}), (\ref{1/8}) can have only a finite number of negative
eigenvalues.
\end{thebibliography}
\end{document}